\documentclass[namedreferences]{SolarPhysics}
\usepackage[optionalrh,solaenum]{spr-sola-addons} 
\usepackage{graphicx}                    
\usepackage{color}                       
\usepackage{url}                         

\begin{document}

\begin{article}

\begin{opening}

\title{Power Spectrum Analysis of Physikalisch-Technische Bundesanstalt Decay-Rate Data: Evidence for Solar Rotational Modulation}

%
\author{P.A.~\surname{Sturrock}$^{1}$\sep
        J.B.~\surname{Buncher}$^{2}$\sep
        E.~\surname{Fischbach}$^{2}$\sep
        J.T.~\surname{Gruenwald}$^{2}$\sep
        D.~\surname{Javorsek}$^{3}$\sep
        J.H.~\surname{Jenkins}$^{4}$\sep
        R.H.~\surname{Lee}$^{5}$\sep
        J.J.~\surname{Mattes}$^{2}$\sep
        J.R.~\surname{Newport}$^{2}$\sep      
       }

%
\runningauthor{P.A.~Sturrock, et al.}
\runningtitle{Power Spectrum Analysis of PTB Decay-Rate Data}

%
  \institute{$^{1}$ Center for Space Science and Astrophysics, Stanford University, Stanford, CA 94305-4060, USA
                     email: \url{sturrock@stanford.edu} \\ 
             $^{2}$ Physics Department, Purdue University, West Lafayette, IN 47907, USA \\
             $^{3}$ 416$^{th}$ Flight Test Squadron, 412th Test Wing, Edwards AFB, CA 93524, USA \\
             $^{4}$ School of Nuclear Engineering, Purdue University, West Lafayette, IN 47907, USA \\
             $^{5}$ Department of Physics, United States Air Force Academy, CO 80920, USA \\ 
             }

\begin{abstract}
Evidence for an anomalous annual periodicity in certain nuclear decay data has led to speculation 
concerning a possible solar influence on nuclear processes. We have recently analyzed data concerning the decay rates 
of $^{36}$Cl and $^{32}$Si, acquired at the Brookhaven National Laboratory (BNL), to search for evidence 
that might be indicative of a process involving solar rotation. Smoothing of the power spectrum by 
weighted-running-mean analysis leads to a significant peak at frequency 11.18 yr$^{-1}$, which is 
lower than the equatorial synodic rotation rates of the convection and radiative zones. This article 
concerns measurements of the decay rates of  $^{226}$Ra acquired at the 
Physikalisch-Technische Bundesanstalt (PTB) in Germany. We find that a similar (but not identical) 
analysis yields a significant peak in the PTB dataset at frequency 11.21 year$^{-1}$, and a peak 
in the BNL dataset at 11.25 yr$^{-1}$. The change in the BNL result is not significant since the 
uncertainties in the BNL and PTB analyses are estimated to be 0.13 year$^{-1}$ and 0.07 year$^{-1}$, 
respectively. Combining the two running means by forming the joint power statistic leads to a highly significant peak 
at frequency 11.23 yr$^{-1}$. We comment briefly on the possible implications of these results for 
solar physics and for particle physics.
\end{abstract}

%

\end{opening}

%
 \section{Introduction} \label{s:Intro} 

For the last one hundred years, scientists have believed that the decay rate of each radioactive element 
is constant, unaffected by any environmental process. In 1930, Rutherford and his colleagues 
wrote ``The rate of transformation of an element has been found to be a constant under all 
conditions'' \cite{Rutherford30} . In the latter half of the 20th century, 
however, there were indications that it is possible to alter the decay rates of certain 
isotopes \cite{Emery72,Hahn76,Dostal77} by physical or chemical processes.

We have recently found strong evidence \cite{Jenkins09b,Fischbach09,Javorsek10} 
for an annual periodicity in decay data acquired at the Brookhaven National 
Laboratory [BNL; \cite{Alburger86}] and at the Physicalisch-Technische 
Bundesandstalt [PTB; \cite{Siegert98}] [although there appears to be a difference 
between early PTB data and more recent data \cite{Schrader92,Schrader10}]. The BNL experiment monitored the decay 
rates of $^{32}$Si and $^{36}$Cl over the interval 1982 to 1989. (Alburger et al. preferred to analyze only data 
from 1982-1986, but we have preferred to analyze the entire data set.) The PTB experiment monitored the 
decay rates of $^{152}$Eu, $^{154}$Eu, and $^{226}$Ra from 1983 to 1998. These analyses, and an apparent change in the 
measured decay rate of $^{54}$Mn during a solar flare in December 2006 \cite{Jenkins09a}, have 
led to the suggestion that some nuclear decay rates are influenced by particles (possibly neutrinos) 
or fields emanating from the Sun \cite{Jenkins09b,Fischbach09}.

As one would expect, the suggestion that nuclear decay rates may be variable has not gone unchallenged:
\begin{enumerate}
\item Semkow et al. (2009) and others have suggested that these fluctuations have their origin in environmental or systematic processes. However, detailed investigation shows that the results of our BNL and PTB analyses cannot be explained by variations of temperature, pressure, humidity, etc. \cite{Jenkins10}.
\item Norman et al. (2009), have re-examined data from several nuclear-decay experiments, finding no evidence for a correlation with Sun-Earth distance. However, our collaboration has re-analyzed the Norman group's data, which Norman and his collaborators generously provided, and we have detected an annual periodicity, small in amplitude but with the same phase as we have found in the BNL and PTB datasets.
\item Cooper (2009) has analyzed data from the power output of the radioisotope thermoelectric generators aboard the Cassini spacecraft, finding no significant deviations from exponential decay, but we find no conflict between Cooper's results and our results \cite{Jenkins10}.
\end{enumerate}

To pursue the question of a possible association between solar processes and variations in nuclear decay rates, we have recently examined the BNL dataset, searching for a periodicity that may be associated with solar rotation \cite{Sturrock10a}--in particular, with rotation rates that show up in low-energy solar neutrino data. We have found that low-energy solar neutrinos, as detected by the Homestake \cite{Davis68,Cleveland98} and GALLEX \cite{Anselmann93,Anselmann95,Hampel96,Hampel99} experiments, exhibit a periodicity at about 12 year$^{-1}$ \cite{Sturrock08,Sturrock09}, which we find also in ACRIM irradiance data \cite{Willson79,Willson01}. Since this frequency is significantly lower than the dominant synodic rotation frequency of the radiative zone, which is about 13 year$^{-1}$ (14 year$^{-1}$ sidereal; \cite{Schou98}), we have suggested that it may be related to the synodic rotation frequency of the solar core (the rotation rate of the core is ill-determined at this time).

In our recent analysis of BNL data, we find a strong peak at 11.18 year$^{-1}$ \cite{Sturrock10a}. The purpose of this article is to carry out a similar analysis of PTB data to see whether these data also show evidence of a periodicity at a frequency lower than the radiative-zone rotation rate.

The preparation of the data and the power spectrum analysis are presented in Section 2, where we focus on a ``search band'' of 10-15 year$^{-1}$. This band extends high enough to cover the equatorial synodic frequencies of the radiative and convection zones \cite{Schou98}, and extends well below the frequency of the strong peak in the BNL power spectrum. In order to assess the probable errors of our frequency estimates, we note that the Fourier 
transform of a pure sinusoidal signal of length $T$ at frequency $ \nu_o $ peaks at $ \nu_o $  and falls to zero at  $ \nu_o \pm \delta\nu $, where $ \delta\nu = 1/T $. Since the durations of the BNL and PTB datasets are 7.80 years and 14.84 years, we adopt the generous error estimates 0.13 year$^{-1}$ and 0.07 year$^{-1}$, respectively. With this guideline, we find a feature in the PTB power spectrum at frequency 11.29 $\pm$ 0.07 year$^{-1}$, close to the frequency of the major peak in the BNL power spectrum. We examine the significance of this periodicity in Section \ref{s:FreqAvgdPS}, using both the standard shuffle test and a variant of this test that we call the ``shake'' test.

Since the power spectrum is more complex than we would expect of a stable, rigid rotator, we adopt in Section \ref{s:FreqAvgdPS} the procedure we used in our analysis of BNL data--that of forming the weighted running mean of the power spectrum, which leads to a prominent feature at frequency 11.25$\pm$0.07 year$^{-1}$. In Section \ref{s:CompComb}, we examine the correlation between the BNL and PTB datasets by forming the ``joint power statistic'' derived from the BNL and PTB weighted-running-mean power spectra. This shows one dominant feature at frequency 11.23$\pm$0.07 year$^{-1}$. We evaluate the significance of this feature by again using the shuffle test. We discuss these results in Section \ref{s:Disc}.

\section{Power Spectrum Analysis} \label{s:PSA}

\begin{figure}[h] 
 \centerline{\includegraphics[width=0.75\textwidth,clip=]{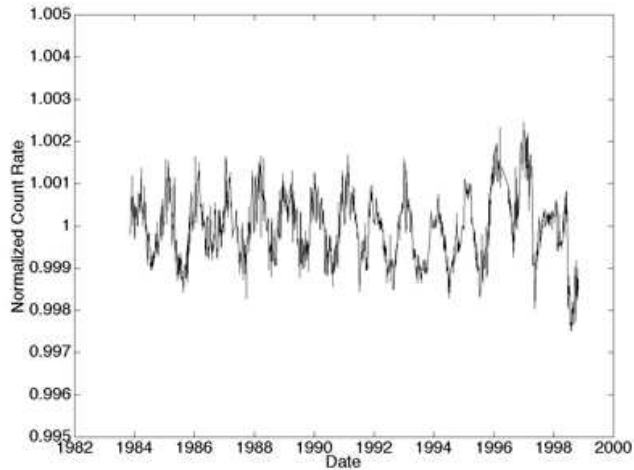}}
 \caption{PTB decay-rate measurements, corrected for mean decay rate and normalized to
mean value unity.}\label{fig:Fig1}
 \end{figure}

The PTB dataset comprises 1966 measurements of the decay rate of $^{226}$Ra in the time interval 1983.855 to 1998.809, where dates are measured in what we call ``neutrino years,'' which have proved useful in analyzing neutrino data. Dates in ``neutrino days'' are counted from January 1, 1970, as day 1, and dates in ``neutrino years'' are given by 1970 + (neutrino days)/365.2564. We have prepared the data for time-series analysis by dividing the decay count rates by the counts expected on the basis of a purely exponential decay, using the mean decay rate determined from the best fit to the data. We also remove a few ``outliers,'' namely datapoints for which the measurements differ from the mean by more than three sigmas. The resulting normalized count rate is shown in Figure \ref{fig:Fig1}, in which the annual variation is obvious.

\begin{figure}[h] 
 \centerline{\includegraphics[width=0.75\textwidth,clip=]{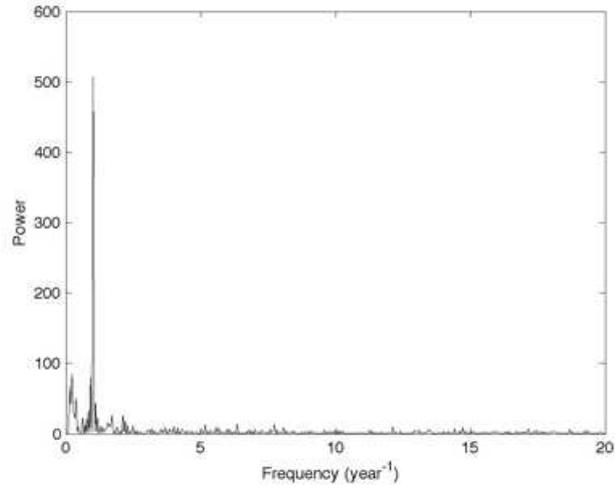}}
 \caption{Power spectrum of PTB data, formed by a likelihood procedure.}\label{fig:Fig2}
 \end{figure}

We have next performed a power-spectrum analysis of the data shown in Figure \ref{fig:Fig1}, in frequency steps of 0.01 year$^{-1}$, using a likelihood procedure \cite{Sturrock05a} that is equivalent to the Lomb-Scargle procedure \cite{Lomb76,Scargle82}. The result is shown in Figure \ref{fig:Fig2}. There is, as expected, a huge peak at 1.00 year$^{-1}$, with power $S=507$, due to the annual modulation which is obvious in Figure \ref{fig:Fig1}. Between 0.01 year$^{-1}$ and 2.11 year$^{-1}$, there are 13 peaks with power 20 or more. We also note that there are strong peaks near 51 and 53 year$^{-1}$ (not shown in this figure), which are obviously due to aliasing of the peak at 1.00 year$^{-1}$ due to a weekly pattern in data acquisition.

Our main goal in this article is to search for features in the power spectrum that might be related to solar rotation. We shall adopt a search band of 10-15 year$^{-1}$ that is wide enough to include periodicities found in previous analyses of related data--neutrino data and irradiance data \cite{Sturrock08,Sturrock09}, and BNL decay data \cite{Sturrock10a}. It is also wide enough to include the band of synodic rotation rates of an equatorial section of the convection and radiative zones \cite{Schou98}. With this goal in mind, it is prudent to remove the influence of the very strong low-frequency modulations, which can cause aliasing and otherwise adversely influence the power spectrum. We find we can suppress most of the low-frequency modulation without appreciably degrading modulation within or above the rotational band by the following procedure.

\begin{figure}[h] 
 \centerline{\includegraphics[width=0.75\textwidth,clip=]{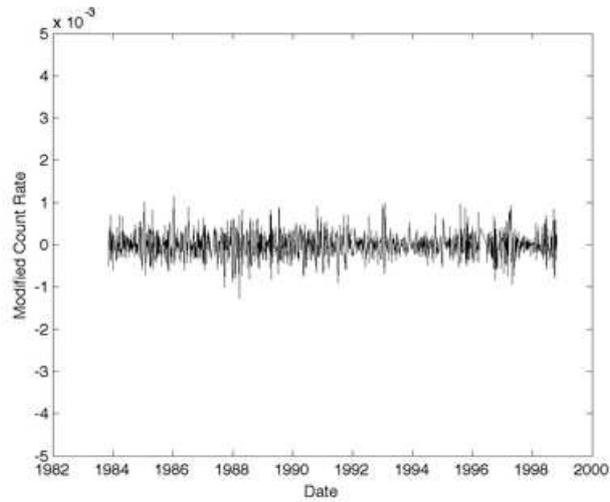}}
 \caption{PTB decay-rate modified count rates measurements.}\label{fig:Fig3}
 \end{figure}

\begin{figure}[h] 
 \centerline{\includegraphics[width=0.75\textwidth,clip=]{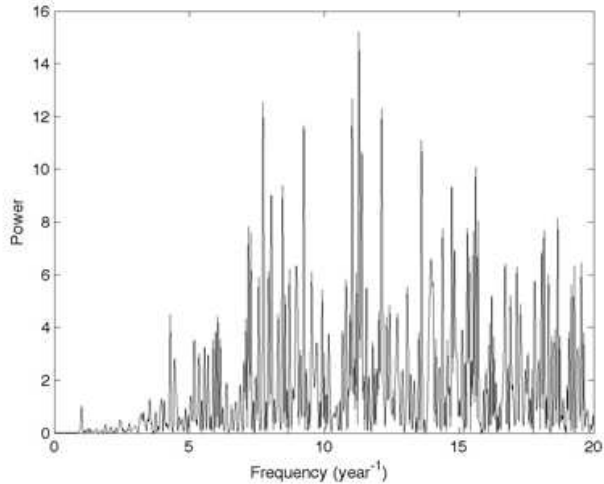}}
 \caption{Power spectrum of modified PTB data, formed by a likelihood procedure. The principal peak is found at 11.29 year$^{-1}$ with power $S = 15.20$.}\label{fig:Fig4}
 \end{figure}

We form 11-point running means of the normalized count rates, and then subtract the running-mean values from the normalized rates. The resulting modified count rates are shown in Figure \ref{fig:Fig3}. A likelihood analysis of this time series yields the power spectrum shown in Figure \ref{fig:Fig4}. The principal feature is at frequency $11.29\pm0.007$ year$^{-1}$ with power $S=15.20$, but there is a cluster of peaks in that vicinity. 

Noting there are 48 peaks in the band 10-15 year$^{-1}$, we find that the standard formula for the false-alarm probability \cite{Scargle82}

\begin{equation}
FAP=1-\left(1-e^{-S}\right) ^{M}\label{eq:Eq01}
\end{equation}

\noindent{}yields the estimate $1.2\times10^{-5}$. However, as we have recently pointed out \cite{Sturrock10b}, if one is over-sampling (as in the present case), one needs to replace $S$ in Equation \ref{eq:Eq01} by $S-1$. With this change, Equation \ref{eq:Eq01} yields the revised estimate $3\times10^{-5}$.

\begin{figure}[h] 
 \centerline{\includegraphics[width=0.75\textwidth,clip=]{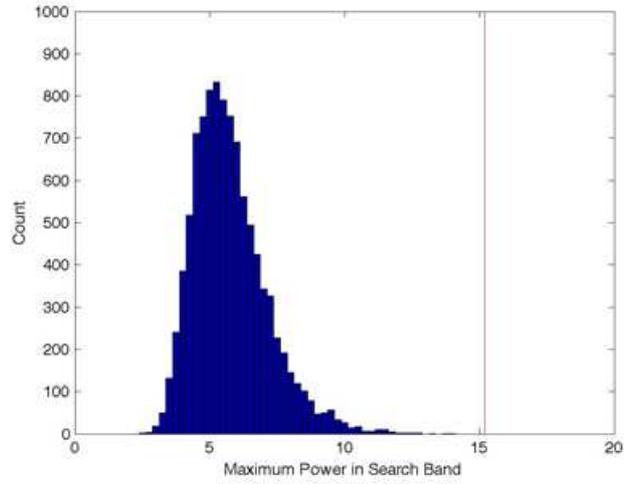}}
 \caption{Histogram display of the result of 10,000 shuffle simulations of the modified PTB dataset. None of the simulations yields as big a power (15.20) in the search band (10-15 year$^{-1}$) as the actual dataset.}\label{fig:Fig5}
 \end{figure}

\begin{figure}[h] 
 \centerline{\includegraphics[width=0.75\textwidth,clip=]{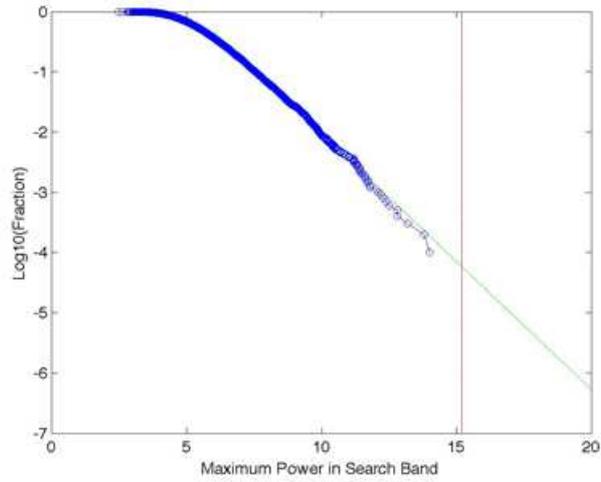}}
 \caption{Logarithmic display of the result of 10,000 shuffle simulations of the modified PTB dataset. A projection of the curve indicates that there is a probability of obtaining by chance as big a power (15.20) in the search band (10-15 year$^{-1}$) as the actual dataset is about $5\times10^{-5}$.}\label{fig:Fig6}
 \end{figure}

In order to obtain a more robust significance estimate, we have carried out the familiar shuffle test \cite{Bahcall91}. We find that, of 10,000 such simulations, none has a power as large as 15.20 in the search band. The distribution is shown as a histogram in Figure \ref{fig:Fig5} and in logarithmic form in Figure \ref{fig:Fig6}. A projection of the distribution shown in Figure \ref{fig:Fig6} indicates that the probability of obtaining by chance a power as large as or larger than 15.20 is only  $5\times10^{-5}$, close to the FAP value.

We have also applied the shake test \cite{Sturrock10a} that involves small random time displacements. The results are very similar to those obtained from the shuffle test.

\section{Frequency-Averaged Power Spectra} \label{s:FreqAvgdPS}

As in our analysis of BNL data, we see from the power spectrum shown in Figure 4 that there is actually a cluster of peaks in the vicinity of 11 year$^{-1}$, indicating that the time series should not be attributed to a single stationary periodic process. The combination of periodicities may be due to the influence of different sources with different characteristic time scales, for example, different rotation rates. For this reason, we adopt a procedure for evaluating a complex of peaks in the search band, rather than focusing on only one peak. This procedure is that of forming a weighted running mean of the power spectrum. 

We denote by $\nu_j$ the sequence of frequencies (with spacing 0.01 year$^{-1}$) at which the power is evaluated, and by $S_j$ the power sequence, where $j=1,\ldots,N_{\nu}$. We then form the sequence

\begin{equation}
\tilde{S}=\sum_{k} W\left( j-k \right) S_k .\label{eq:Eq02}
\end{equation}

\noindent{}We have adopted

\begin{equation}
W \left( j \right) = C \cos \left( \frac{\pi}{2} \frac{j}{m} \right), j=-m\ldots m,\label{eq:Eq03}
\end{equation}

\noindent{}in which $C$ is chosen so that the sum of the weighting terms $W$ is unity:

\begin{equation}
C=\left[ \sum_{-m}^{m} \cos \left( \frac{\pi}{2} \frac{j}{m} \right) \right] ^{-1}.\label{eq:Eq04}
\end{equation}

\noindent{}We have adopted $m = 50$, which is equivalent to forming running means of the power over intervals of width 1 year$^{-1}$. However, we find that

\begin{equation}
\left( \sum_{-m}^{m} W\left( j \right) j^{2} \right)^{1/2}=0.435m,\label{eq:Eq05}
\end{equation}

\noindent{}so that, for $m = 50$, the rms frequency deviation is only 0.22 year$^{-1}$.

\begin{figure}[h] 
 \centerline{\includegraphics[width=0.75\textwidth,clip=]{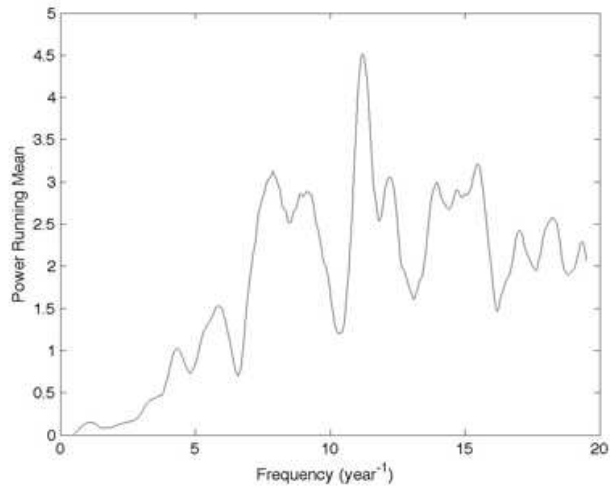}}
 \caption{Plot of the 101-point weighted running means formed from the power spectrum shown in Figure \ref{fig:Fig4}. The peak occurs at 11.21 year$^{-1}$ with weighted-running-mean power 4.51.}\label{fig:Fig7}
 \end{figure}

We show in Figure \ref{fig:Fig7} the result of applying this smoothing operation to the power spectrum shown in Figure \ref{fig:Fig4} . We find that the smoothed power spectrum has its principal feature at 11.21$\pm$0.07 year$^{-1}$  with peak power 4.51. 

\begin{figure}[h] 
 \centerline{\includegraphics[width=0.75\textwidth,clip=]{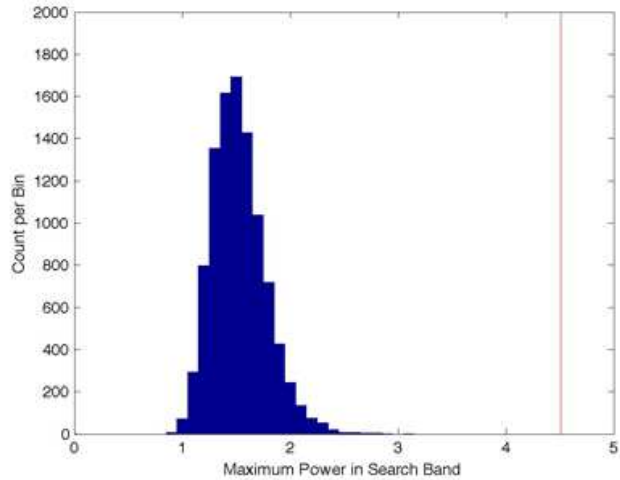}}
 \caption{Histogram of the weighted-running-mean powers computed for 10,000 shuffle simulations of the modified PTB data. None is as large as the actual value (4.51).}\label{fig:Fig8}
 \end{figure}

Since the running means are not distributed exponentially, it is necessary to carry out the shuffle test or a similar test to obtain a significance estimate. We again use the shuffle test, re-computing the power spectrum and then the running mean of the power spectrum many times, keeping the original times and the original count rates, but randomly sorting one of these datasets. We have carried out this procedure 10,000 times, determining the maximum weighted running mean of the power in the rotational search band for each simulation. The histogram formed from these maxima is shown in Figure \ref{fig:Fig8}. We see that none of the random simulations yields as large a value of the running mean of the power in the search band as the actual value (4.51).

\begin{figure}[h] 
 \centerline{\includegraphics[width=0.75\textwidth,clip=]{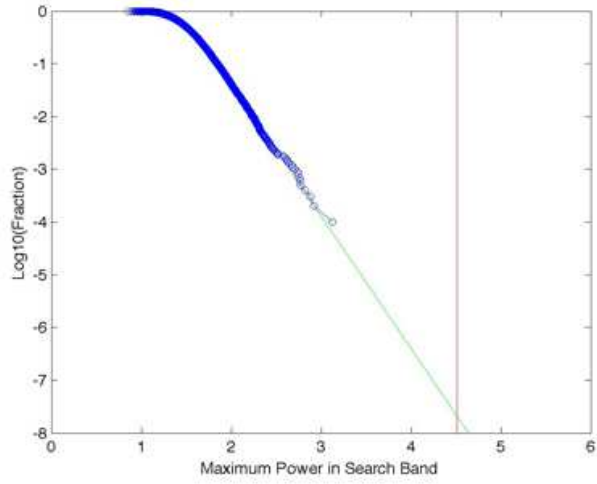}}
 \caption{Logarithmic plot of the weighted-running-mean powers computed for 10,000 shuffle simulations of the modified PTB data. A projection indicates that there is a probability of only about $2\times10^{-8}$ of obtaining by chance a values as large as the actual value (4.51).}\label{fig:Fig9}
 \end{figure}

These results are shown in a logarithmic display in Figure \ref{fig:Fig9} , which also shows an extension of an empirical curve derived from a projection of the last 1,000 points. This projection leads us to a $p$-value of 10$^{-7.7}$, implying that we could expect to obtain a peak as large or larger than the actual peak only once in about $2\times10^{8}$ simulations.

We have also examined the running-mean power spectrum by means of the shake test, again carrying out 10,000 simulations. We find that the results are very similar to those obtained by the shuffle test.

\section{Comparison and Combination of BNL and PTB Data} \label{s:CompComb}

\begin{figure}[h] 
 \centerline{\includegraphics[width=0.75\textwidth,clip=]{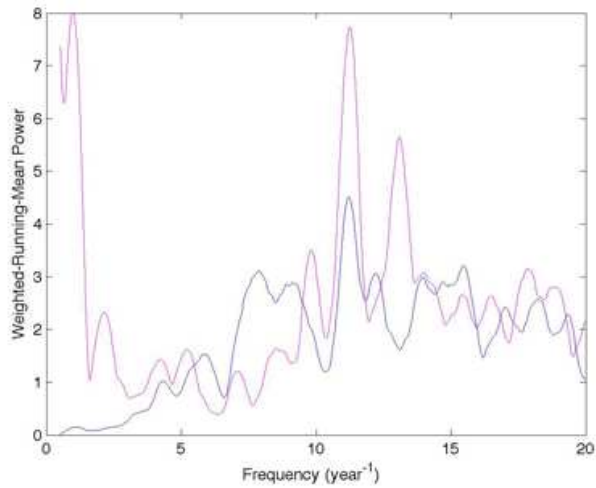}}
 \caption{Plots of the 101-point weighted running means formed from the PTB power spectrum (blue) and from the BNL power spectrum (magenta). The two peaks are found at 11.21 year$^{-1}$ and at 11.25 year$^{-1}$, respectively.}\label{fig:Fig10}
 \end{figure}

We now derive weighted-running-means of the BNL power spectrum. To be consistent with our analysis of PTB, we here again adopt $m = 50$. We show the result, together with the PTB running-mean power spectrum for comparison, in Figure \ref{fig:Fig10}. We find that the BNL curve has a peak at $11.25$ year$^{-1}$ with uncertainty $0.13$ year$^{-1}$  and running-mean power 7.74. This is indistinguishable from the peak in the PTB curve at $11.21$ year$^{-1}$  that has uncertainty  $0.07$ year$^{-1}$. Taking into account the uncertainty of  $0.13$ year$^{-1}$, it is also indistinguishable from our earlier estimate of $11.18$ year$^{-1}$ \cite{Sturrock10a}. 

It is useful to examine the degree of correlation between the BNL and PTB power spectra. A convenient procedure is to form the ``joint power statistic'' \cite{Sturrock05b}. For the present situation, in which we are considering only two power spectra, we first form the geometric mean:

\begin{equation}
X=\left( S_1 S_2 \right)^{1/2}. \label{eq:Eq06}
\end{equation}

\noindent{}The joint power statistic is then defined by

\begin{equation}
J_{2}=-\ln \left( 2XK_{1}\left( 2X \right) \right) \label{eq:Eq07}
\end{equation}

\noindent{}where $K_1$ is the Bessel function of the second kind. This statistic has the useful property that if each power is distributed exponentially, so that

\begin{equation}
P\left( S\right) dS = e^{-S}dS, \label{eq:Eq08}
\end{equation}

\noindent{}then $J$ satisfies the same distribution. We find that the following simple expression offers a close approximation to that of Equation \ref{eq:Eq07}:

\begin{equation}
J_{2A}=\frac{1.943X^{2}}{0.650+X}. \label{eq:Eq09}
\end{equation}

We now compute the joint power statistic for the running-means that we have calculated for the BNL and PTB power spectra. This offers a useful comparison with the running-mean of each power spectrum. We show the result of this calculation in Figure 11. We find a prominent feature at frequency $11.23\pm0.07$ year$^{-1}$  with $J$ = 10.34.

\begin{figure}[h] 
 \centerline{\includegraphics[width=0.75\textwidth,clip=]{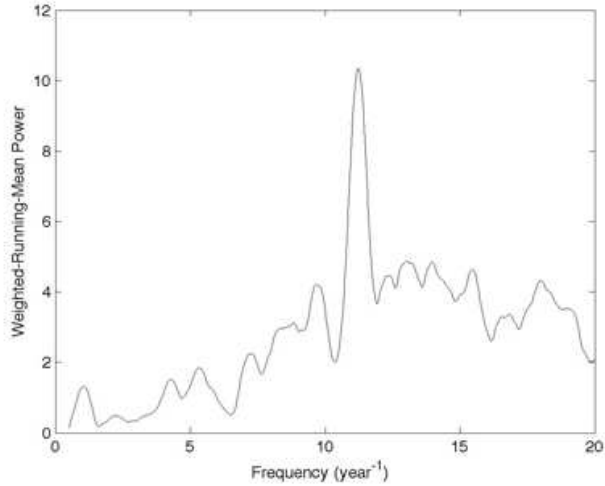}}
 \caption{The joint power spectrum formed from the weighted running-mean BNL and PTB power spectra. The peak is found at 11.23 year$^{-1}$ with $J=$10.34.}\label{fig:Fig11}
 \end{figure}

We again use the shuffle test, computing 10,000 simulations of the calculations that led to Figure \ref{fig:Fig11}.   For each simulation, we retain the actual measurements and the actual times for each dataset, but re-assign them randomly. For each simulation, we compute the two power spectra, form the running means, and then find the maximum value of the joint power statistic in the search band 10-15 year$^{-1}$.

\begin{figure}[h] 
 \centerline{\includegraphics[width=0.75\textwidth,clip=]{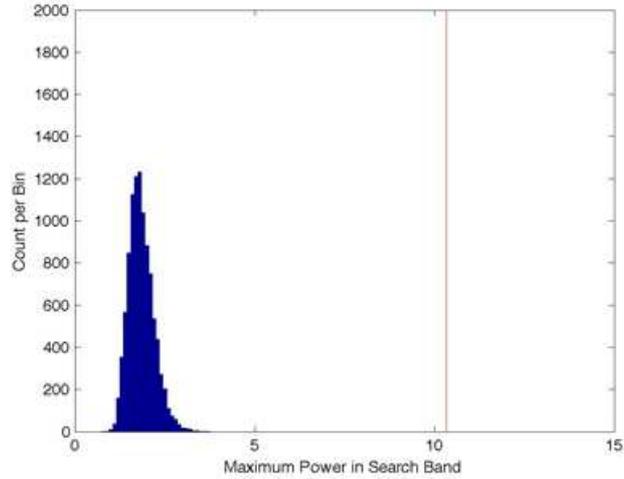}}
 \caption{Histogram of the weighted-running-mean powers computed for 10,000 shuffle simulations of the joint power statistic formed from the BNL and PTB running-mean power spectra. None is as large as the actual value (10.34).}\label{fig:Fig12}
 \end{figure}

\begin{figure}[h] 
 \centerline{\includegraphics[width=0.75\textwidth,clip=]{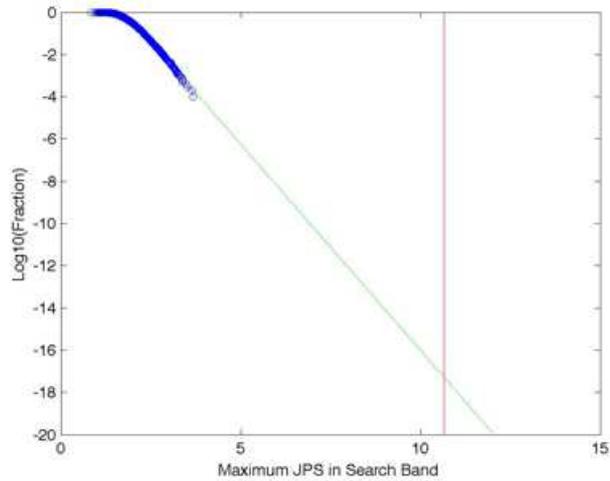}}
 \caption{Logarithmic plot of the joint power statistic computed for 10,000 shuffle simulations of the joint power statistic formed from the BNL and PTB weighted-running- mean power spectra.  A projection indicates that there is a probability of only about 10$^{-17}$ of obtaining by chance a value as large as the actual value (10.34).}\label{fig:Fig13}
 \end{figure}

We show the distribution of these maximum values in histogram form in Figure \ref{fig:Fig12} , and in logarithmic form in Figure \ref{fig:Fig13}. The maximum value found in these 10,000 simulations is 3.67, whereas the value obtained from actual data is 10.34. Figure \ref{fig:Fig13} shows an extension of an empirical curve derived from a projection of the last 1,000 points. This projection leads us to a $p$-value of 10$^{-17.2}$, implying that we could expect to obtain a peak as large or larger than the actual peak only once in about 10$^{17}$ simulations.

\section{Discussion} \label{s:Disc}

One must be cautious about accepting a p-value as small as 10$^{-17.2}$ at face value. Nevertheless, there is clearly strong evidence that both BNL and PTB decay-rate measurements are influenced by a common periodicity with a frequency of about 11.25 year$^{-1}$. This is too low a frequency to reflect a rotational modulation that originates in the radiative zone or a near-equatorial region of the convection zone. This leaves only the innermost region of the Sun, with normalized radius of approximately 0.2 \cite{Garcia07}, for which we do not yet have reliable rotation measurements.

However, the possibility that decay-rate variations are due to solar neutrinos provides an independent reason for focusing attention on the core, since that is the location of nuclear reactions that produce neutrinos. It is also relevant to recall that we have previously found evidence that low-energy solar neutrinos exhibit a periodicity at or near 11.85 year$^{-1}$ \cite{Sturrock08,Sturrock09}, lower than the estimated synodic rotation rate of the radiative zone. These considerations suggest that the sub-radiative-zone region of the Sun is not in rigid rotation. If this is so, we must expect that neutrinos of different energies will have different periodicities. As a consequence, it is possible that the decay-rates of different nuclei will be different.

A possible scenario is that the central part of the core has a synodic rotation rate of about 11.25 year$^{-1}$, corresponding to a sidereal rotation rate of 12.25 year$^{-1}$. There must then be a radial gradient of rotation rate between the core and the radiative zone, that is believed to have a sidereal rotation rate of about 13.60 year$^{-1}$ \cite{Schou98}. Such an intermediate region would constitute an ``inner tachocline,'' which might lead to dynamo action, similar to the dynamo action attributed to the outer tachocline \cite{Krause92}.  The periodicity of 11.85 year$^{-1}$, detected in low-energy-neutrino data and total solar irradiance data \cite{Sturrock08,Sturrock09}, corresponding to a sidereal rotation rate of 12.85 year$^{-1}$, would originate somewhere in the inner tachocline.

A rotating core can lead to a corresponding periodicity only if the core is asymmetric. Hence it seems plausible to attribute periodicities in neutrino, irradiance, and decay-rate data in the range 11-12 year$^{-1}$ to processes involving the rotation of a solar core that is not cylindrically symmetric. This raises the possibility that the core may also exhibit north-south asymmetry. In this connection, it is worth noting that, since the Sun's axis is inclined with respect to the normal to the ecliptic, a north-south asymmetry might contribute to the annual variation found in decay rates \cite{Jenkins09b,Fischbach09,Javorsek10}. This effect, in association with an annual variation due to the eccentricity of the Earth's orbit, may explain why the phases of the annual variations in decay data do not agree with what would be expected on the basis of a purely orbital effect. This issue will be discussed in detail in a subsequent article. 

These results raise questions concerning not only solar physics but also particle physics. They also point to the need for new experiments that monitor the decay rates of a variety of nuclei with high cadence and high accuracy.

%

%

%

%
 \begin{acks}
The authors are indebted to D. Alburger and G. Harbottle for supplying us with raw data from the BNL experiment, to H. Schrader for supplying the raw data from the PTB experiment, and to the referee who made a number of suggestions that significantly improved the article. The work of PAS was supported in part by the National Science Foundation through grant AST-0097128, and that of EF was supported in part by U.S. DOE contract No. DE-AC02-76ER071428.
 \end{acks}

%
%
%
%
%
%

\end{article} 
\end{document}